\documentclass[aps,pra,superscriptaddress,reprint,floatfix,longbibliography,showpacs,amsfonts,amsmath,amssymb]{revtex4-1}
\usepackage{graphicx}
\usepackage{hyperref}
\usepackage{xr-hyper}
\usepackage{upgreek}
\usepackage{epstopdf}
\usepackage[note-name=, use-sort-key = false]{notes2bib}
\usepackage{color}
\usepackage{comment}

\begin{document}
\title{Optical conductivity of the type-II Weyl semimetal WTe$_2$ under pressure}

\author{M. Krottenm\"uller}
\affiliation{Experimentalphysik II, Institute of Physics, University of Augsburg, 86159 Augsburg, Germany}
\author{J. Ebad-Allah}
\affiliation{Experimentalphysik II, Institute of Physics, University of Augsburg, 86159 Augsburg, Germany}
\affiliation{Department of Physics, Tanta University, 31527 Tanta, Egypt}
\author{V. S\"uss}
\affiliation{Max Planck Institute for Chemical Physics of Solids, N\"othnitzer Strasse 40, 01187 Dresden, Germany}
\author{C. Felser}
\affiliation{Max Planck Institute for Chemical Physics of Solids, N\"othnitzer Strasse 40, 01187 Dresden, Germany}
\author{C. A. Kuntscher}
\email{christine.kuntscher@physik.uni-augsburg.de}
\affiliation{Experimentalphysik II, Institute of Physics, University of Augsburg, 86159 Augsburg, Germany}

\begin{abstract}
Tungsten ditelluride WTe$_2$ is a type-II Weyl semimetal with electronic properties highly sensitive to external pressure, as demonstrated by the
superconductivity emerging under pressure.
Here, we study the optical conductivity of the type-II Weyl semimetal WTe$_2$ under external pressure at room temperature.
With increasing pressure, a pronounced spectral weight transfer occurs from the high-energy to the low-energy interband transitions, with drastic changes in the profile of the optical conductivity spectrum indicating a high sensitivity of the electronic band structure to external pressure.
The detailed analysis of the pressure-dependent optical conductivity furthermore reveals anomalies at the pressures $\sim$2 and $\sim$4.5~GPa, where an electronic and a structural phase transition, respectively, were reported in the literature.
\end{abstract}

\maketitle

\section{Introduction}
The layered transition-metal dichalcogenide WTe$_2$ is currently extensively studied \cite{Pan.2018,Tian.2018,Das.2019}, as it is an inversion-symmetry-breaking type-II Weyl semimetal \cite{Soluyanov.2015,Lin.2017,Li.2017} with an extremely large nonsaturating magnetoresistance (MR) \cite{Ali.2014,Zhu.2015,Onishi.2018,Jha.2018} and other outstanding properties, such as a non-linear anomalous Hall effect in few-layer samples, despite being non-magnetic \cite{Kang.2019,Shvetsov.2019}, room-temperature ferroelectricity \cite{Sharma.2019}, unconventional Nernst effect \cite{Rana.2018}, and pressure-induced superconductivity \cite{Kang.2015,Pan.2015}. An ultrafast symmetry switch has been realized in WTe$_2$ \cite{Sie.2019}, exploiting structural changes induced by teraherz radiation and the accompanying changes in the topological Weyl state.
The large MR in WTe$_2$ is usually explained by a perfect compensation of electron and hole carriers \cite{Ali.2014}, which is supported by  theoretical calculations \cite{Lv.2015} as well as ARPES and transport experiments \cite{Pletikosic.2014, Cai.2015, Pan.2017,Wang.2019}. However, the results of Hall effect measurements questioned this scenario \cite{Luo.2015,Wang.2016} and alternative explanations such as strong spin-orbit coupling and forbidden backscatterings due to the spin texture were proposed \cite{Rhodes.2015,Jiang.2015}. Therefore, knowledge of the properties of the Fermi surface of WTe$_2$ is highly desirable. According to angle-dependent transport measurements, the electronic properties of WTe$_2$ are rather three-dimensional, and the temperature dependence of the magnetoresistance follows the temperature dependence of the mass anisotropy and thus the anisotropy of the Fermi surface \cite{Thoutam.2015}. Furthermore, nonlinear MR and its temperature behavior were explained by the temperature-induced changes in Fermi surface convexity \cite{He.2019}.

Shubnikov-de Haas experiments under pressure \cite{Cai.2015} not only favor the scenario of two electron and two hole pockets of similar size, but also highlight the drastic pressure dependence of the Fermi surface of WTe$_2$. Accordingly, with increasing pressure, a strong increase of the size of the Fermi surface is observed, as well as a change in its topology, namely two pockets disappear around 1~GPa. Interestingly, two other studies on WTe$_2$ discovered superconductivity under pressure with a maximum T$_c\approx$7~K, although with a discrepancy in the pressure onset of the transition (10.5, 2.5~GPa)\cite{Kang.2015,Pan.2015} coinciding with the suppression of the large MR effect. It was furthermore found that WTe$_2$ undergoes a pressure-induced structural phase transition from the ambient-pressure orthorhombic T$_\mathrm{d}$ to the monoclinic T' phase at $\sim$8~GPa \cite{Zhou.2016,Lu.2016,Xia.2017}. Thus, it was suggested that the structural phase transition separates the large MR state from the superconducting state \cite{Lu.2016,Xia.2017}, in contradiction to the findings in Ref.\ \cite{Zhou.2016}, where both polytypes show superconductivity. Furthermore, electronic band structure calculations under pressure show an anisotropic Fermi surface at high pressure \cite{Lu.2016}, whereas recent angle-dependent measurements of the upper critical field reveal a nearly isotropic superconductivity at $\sim$10~GPa \cite{Chan.2017}, not compatible with the calculated Fermi surface anisotropy.

Optical spectroscopy is a powerful technique for probing the charge dynamics in a material with a high energy resolution. In particular, materials with nontrivial topology show a characteristic frequency dependence of their optical conductivity. In particular, 3D Dirac and type I-Weyl semimetals show a frequency-linear behavior due to interband transitions between the linearly dispersing nontrivial bands \cite{Tabert.2016b}, like it was detected, e.g., for TaAs \cite{Xu.2016}. In comparison, the optical conductivity of type-II Weyl semimetals with tilted cones contains two regions with quasi-linear behavior, where the change of slope is a measure for the tilting of the Weyl cones \cite{Carbotte.2016}. Linear extrapolation of the higher-energy region gives a finite conductivity value at zero frequency.

The optical conductivity of WTe$_2$ at ambient pressure \cite{Homes.2015,Frenzel.2017,Kimura.2019} contains two Drude components whose spectral weight and scattering rate showed a markedly different temperature behavior. One Drude term was associated with trivial, semimetallic electron and hole bands, while the other one with Dirac carriers at Weyl points \cite{Kimura.2019}. WTe$_2$ also shows a frequency-linear optical conductivity typical for Dirac and Weyl semimetals; however, according to \textit{ab initio} calculations, in WTe$_2$ the linear behavior of the optical conductivity stems from a sum of many transitions involving trivial bands \cite{Frenzel.2017}, and besides, the two distinct regions with quasi-linear behavior characteristic for type-II Weyl semimetals are lacking.

On the other hand, to our knowledge, such optical signatures for type-II Weyl semimetals could not be unambiguously proven to exist in any candidate material up to now. Although two linear slopes were found in the optical conductivity of YbMnBi$_2$ via reflectivity measurements, both quasi-linear behaviors extrapolate through the origin \cite{Chinotti.2016}. Moreover, it was found that a feature which might be attributed to a van Hove singularity in a simplified Weyl semimetal bandstructure, can more likely be described by certain non-idealities in the present material \cite{Chaudhuri.2017}. Two linear regimes separated by a kink were also observed in the optical conductivity of elemental tellurium under pressure \cite{Rodriguez.2020}. But by band-resolved DFT calculations it was shown that the origin of these features is not due to a Weyl type-II bandstructure. In a recent publication \cite{Mardele.2020}, the authors show the optical conductivity of TaIrTe$_4$ indeed consisting of two linear slopes separated by a kink and interpret these findings in terms of a tilted Weyl dispersion, but other origins of this behaviour could not be excluded.

In order to obtain more detailed information on the electronic properties of the type-II Weyl semimetal WTe$_2$ under pressure, we studied the optical conductivity of WTe$_2$ for pressures up to 8~GPa.
Upon pressure application, apart from the increasing metallicity due to the increase in the Fermi pockets, we observe a large spectral weight transfer from high to low energies. Furthermore, we find indications for two phase transitions, at $\sim$2 and $\sim$4.5~GPa, presumably of electronic and structural type, respectively. Our findings confirm the strong influence of pressure on the bandstructure and thus the peculiar electronic properties of WTe$_2$.

\section{SAMPLE PREPARATION AND EXPERIMENTAL DETAILS}

Single crystals were grown by chemical vapor transport of polycrystalline WTe$_2$ with TeCl$_4$ (Aldrich, 99\%) as a transport additive, the evacuated silica ampoule heated in a two-zone-furnace in a temperature gradient from 900$^\circ$C (T2) to 800$^\circ$C  (T1) for several days. After reaction, the ampoule was removed from the furnace and quenched in water. The 0.5-1 mm size plate-like crystals were characterized by powder x-ray diffraction (XRD). At ambient pressure, WTe$_2$ crystallizes in the noncentrosymmetric, orthorhombic T$_\mathrm{d}$ phase with distorted Te octahedra and zigzag chains of W atoms along the $a$ axis \cite{Brown.1966,Dawson.1987}, in contrast to the monoclinic 1T' structure and the undistorted trigonal prismatic 2H structure.

Pressure-dependent infrared reflectivity measurements were carried out at room temperature in the frequency range 300-16000~cm$^{-1}$ using a Bruker Vertex v80 Fourier transform infrared spectrometer coupled to a Bruker Hyperion infrared microscope. For the measurements, freshly cleaved crystals were placed in the hole of CuBe gaskets inside a screw-driven diamond-anvil cell. Finely ground CsI powder served as quasi-hydrostatic pressure transmitting medium (PTM). The pressure inside the DAC was determined \textit{in situ} using the ruby luminescence technique \cite{Mao.1986,Syassen.2008}. The reflectivity spectra at the sample-diamond interface $R_\mathrm{s-d}(\omega)$ were obtained according to $R_\mathrm{s-d}(\omega)=R_\mathrm{s-gasket}(\omega)\times I_\mathrm{s}(\omega)/I_\mathrm{gasket}(\omega)$, where $I_\mathrm{s}(\omega)$ is the intensity of the radiation reflected at the sample-diamond interface, $I_\mathrm{gasket}$ is the intensity reflected from the CuBe gasket-diamond interface, and $R_\mathrm{s-gasket}(\omega)$ is the reflectivity of the gasket material for the diamond interface. The frequency range of the diamond multi-phonon absorptions 1800-2700~cm$^{-1}$ was interpolated based on a Drude-Lorentz fitting of the reflectivity data.

The $R_\mathrm{s-d}(\omega)$ spectra were treated via Kramers Kronig (KK) analysis taking into account the sample-diamond interface \cite{Pashkin.2006} in order to obtain the complex optical conductivity $\hat{\sigma}(\omega)=\sigma_1(\omega)+\mathrm{i}\sigma_2(\omega)$ and the complex dielectric function $\hat{\epsilon}(\omega)=\epsilon_1(\omega)+\mathrm{i}\epsilon_2(\omega)$. To this end, the low-frequency extrapolations of the reflectivity data were based on Drude-Lorentz fitting, whereas for the high energy extrapolation ambient-pressure x-ray atomic scattering functions were utilized \cite{Tanner.2015}, adjusted for the sample-diamond interface.

\section{RESULTS}

\begin{figure}[t]
\includegraphics[width=0.5\textwidth]{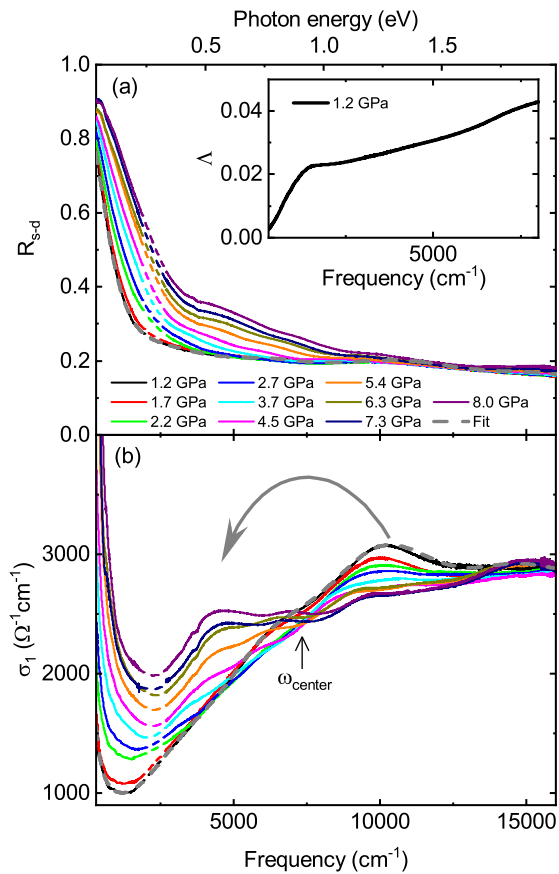}
\caption{(a) Reflectivity spectrum $R_\mathrm{s-d}(\omega)$ of WTe$_2$ for all measured pressures, together with the Drude-Lorentz fit for the spectrum at 1.2~GPa (gray dashed line).  Inset: Loss function $\Lambda$ at the lowest pressure, showing no clear plasmon peak. (b) Real part of the optical conductivity $\sigma_1(\omega)$ of WTe$_2$ for all measured pressures. The gray arrow illustrates the pressure-induced spectral weight transfer from $\omega_\mathrm{high}$ to $\omega_\mathrm{low}$, as explained in the text. The gray dashed line marks the fit of the optical conductivity spectrum at 1.2~GPa.}\label{figure1}
\end{figure}

Pressure-dependent reflectivity $R_\mathrm{s-d}(\omega)$ spectra of WTe$_2$ are depicted in Fig.~\ref{figure1}(a). At the lowest pressure (1.2~GPa), the reflectivity drops quite sharply from $\sim$0.75 at the lowest measured frequency to $\sim$0.3 at 1500~cm$^{-1}$. Above 1500~cm$^{-1}$, the reflectivity decreases monotonically to the value $\sim$0.18 at the highest measured frequency (16000~cm$^{-1}$). The features at intermediate frequencies signal the presence of several interband transitions. The reflectivity drop in the low-energy range of the spectrum corresponds to a plasma edge, which sharpens during cooling down \cite{Homes.2015}. Since the plasma edge is rather smeared out at room temperature and under pressure, no clear plasmon peak appears in the loss function $\Lambda$, defined as $\Lambda = $-Im(1/${\hat{\epsilon}}$) (see inset of Fig.~\ref{figure1}(b)). Instead, the loss function increases strongly up to $\sim$1600~cm$^{-1}$, reaching a plateau, and then increases monotonically in a slower fashion.

With increasing pressure, the overall reflectivity increases, indicating an increasing metallicity as it is expected for semimetallic compounds under pressure. The reflectivity increase is stronger at low frequencies compared to high frequencies.
The plasma edge in the reflectivity spectrum shifts to higher energies with increasing pressure. Above $\approx$4.5~GPa, a shoulder appears at $\approx$4500~cm$^{-1}$.

The real part of the optical conductivity $\sigma_1$, as plotted in Fig.~\ref{figure1}(b), allows for a more detailed view on the optical properties of WTe$_2$ under pressure. At 1.2~GPa, one can see the tail of a small Drude contribution, consistent with the semimetallic nature of WTe$_2$. The value of the optical conductivity at the lowest measured frequency matches well with the dc conductivity of  $\sim$1050~$\Omega^{-1}$cm$^{-1}$ \cite{Jana.2015}. Above 1500~cm$^{-1}$ the optical conductivity increases almost linearly, up to a maximum close to 10000~cm$^{-1}$.
The linear behavior was identified to stem from the sum of many interband transitions involving trivial bands \cite{Frenzel.2017}, and should not be ascribed to interband transitions close to the Weyl points with a characteristic $\omega$-linear conductivity \cite{Ashby.2014,Tabert.2016b,Carbotte.2016}. Above 10000~cm$^{-1}$, the optical conductivity slightly decreases first and then remains constant. Overall, the optical conductivity at 1.2~GPa is consistent with the published ambient-pressure data \cite{Homes.2015,Frenzel.2017,Kimura.2019}\footnote{We point out that the features in our optical conductivity spectra are rather broad compared to literature data, since the measurements were carried out under external pressure at room temperature.}.
The interband transitions in WTe$_2$ occur between several electronic bands with W-5d and Te-5p character in the vicinity of
the Fermi energy E$_F$, including electron and hole bands along the $\Gamma$-$X$ direction in the Brillouin zone \cite{Homes.2015,Pan.2015}.
Furthermore, based on the comparison between the experimental and calculated conductivity spectra, weak electron correlations were inferred \cite{Kimura.2019}.


For a quantitative analysis, we carried out a simultaneous fitting of the reflectivity $R_\mathrm{s-d}(\omega)$ and optical conductivity $\sigma_1(\omega)$ spectra based on a Drude-Lorentz model for the dielectric function $\hat{\epsilon}(\omega)$, containing one Drude contribution and several Lorentz contributions, according to:
\begin{equation}  \label{equ:Drude-Lorentz}
\hat{\epsilon}(\omega)=\epsilon_{\infty}-\frac{\omega_\mathrm{pl,D}^2}{\omega^2+i\Gamma_D\omega}+\sum_{j}\frac{\Omega_j^2}{\omega_{0,j}^2-\omega^2-i\Gamma_\mathrm{j}\omega} \quad,
\end{equation}
with the  frequency $\omega_\mathrm{pl,D}$ and scattering rate $\Gamma_D$ of the Drude contribution, the resonance frequency $\omega_{0,\mathrm{j}}$, the oscillator strength $\Omega_j$ and the full width at half maximum $\Gamma_\mathrm{j}$ of the j-th Lorentz oscillator, and the real part of the dielectric function at high frequency $\epsilon_{\infty}$.
Despite two Drude components being considered in optical conductivity spectra of WTe$_2$ \cite{Homes.2015,Frenzel.2017,Kimura.2019}, we included only one Drude term in our analysis, since our measured range starts above 300~cm$^{-1}$ and the scattering rate of the second Drude term is much smaller \footnote{The small scattering rate and its temperature dependence as well as the rapid increase of the effective electron number of the second Drude term are concluded to be due to Weyl-type carriers \cite{Kimura.2019}}. Besides the Drude term, we had to insert a low-energy Lorentzian contribution L1 at 850~cm$^{-1}$ to adequately describe the low-energy optical conductivity, consistent with recent observations of low-energy excitations in the sister compound MoTe$_2$ \cite{SantosCottin.2019}. We show the complete fitting with its components for the 1.2~GPa spectrum in Fig.~\ref{figure2}(a) (for more information about the fittings see Fig.~\ref{suppl2} and \ref{suppl3} in the appendix).
According to Ref.\ \cite{Homes.2015} the L1 contribution stems from direct transitions between the electronic bands along $\Gamma$-$X$ associated with the hole and electron pockets. It is difficult to attribute the higher-energy Lorentz contributions to specific electronic transitions, since there are many electronic bands in the energy range E$_\mathrm{F}\pm$1~eV \cite{Homes.2015}.

\begin{figure}[t]
\includegraphics[width=0.5\textwidth]{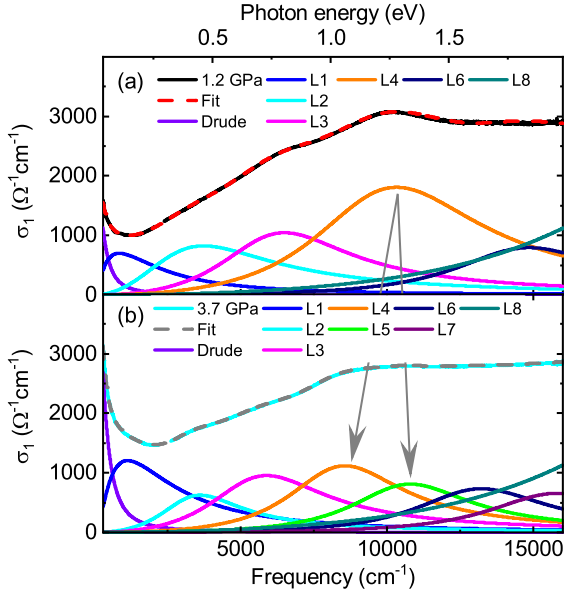}
\caption{Optical conductivity spectr together with the fits and contributions for (a) 1.2~GPa and (b) 3.7~GPa, illustrating the different models needed to describe the data below and above 2~GPa, respectively. The two grey arrows indicate the splitting of the Lorentz contribution L4 into two contributions L4 and L5 above 2~GPa.}
\label{figure2}
\end{figure}

Importantly, in the region 800-7500~cm$^{-1}$ the optical conductivity drastically increases with increasing pressure, while it decreases between 7500 and 12500~cm$^{-1}$ [see Fig.~\ref{figure1}(b)]. Above 12500~cm$^{-1}$, $\sigma_1$ is approximately pressure independent. We also note that at 8~GPa the profile of the optical conductivity $\sigma_1$ has markedly changed as compared to ambient pressure, with an almost constant behavior in a broad energy range from 4000 to 16000~cm$^{-1}$ [see Fig.~\ref{figure1}(b)].
Since $\sigma_1$ is related to the joint density of states JDOS according to $\sigma_1(\omega)\propto \mathrm{JDOS}(\omega)/\omega$ assuming frequency-independent transition matrix elements, the observed pressure-induced changes in $\sigma_1$ indicate drastic changes in the electronic band structure and cannot be explained by a mere shifting of bands. Overall, one observes a pronounced pressure-induced spectral weight transfer in the optical conductivity spectrum from high to low frequencies around the center frequency $\omega_\mathrm{center}=$7500~cm$^{-1}$, approximately forming an isosbestic (equal-absorption) point.
Isosbestic points \cite{Vollhardt.1997,Greger.2013} have also been found in the optical conductivity spectrum of stongly correlated materials, like the cuprates La$_{2-x}$Sr$_x$CuO$_4$ \cite{Uchida.1991} and the pyrochlore-type molybdates \cite{Kezsmarki.2004}. Here, they were discussed in terms of a spectral weight transfer between intraband and interband transitions in the context of a Mott-type metal-insulator-transition within the Hubbard model. However, strongly correlated electron physics is not relevant for WTe$_2$.

\begin{figure}[t]
\includegraphics[width=0.45\textwidth]{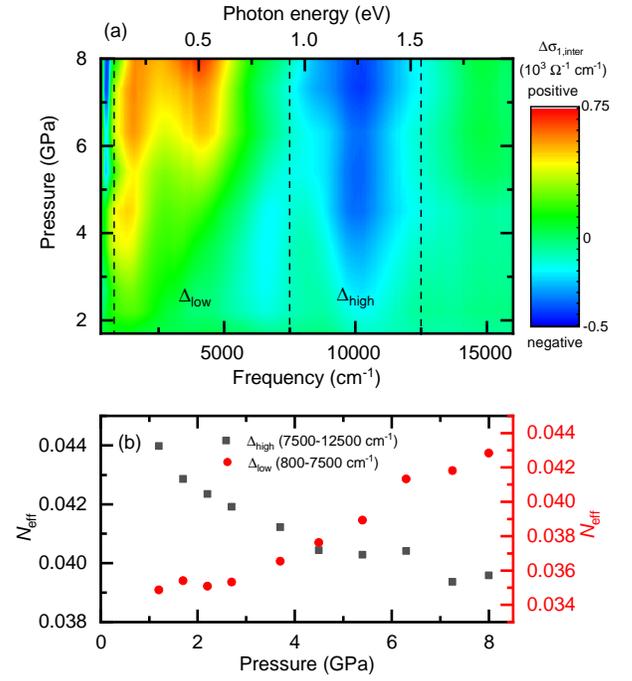}
\caption{(a) Contour plot of the difference spectra $\Delta\sigma_{1,\mathrm{inter}}$ as defined in the text. The vertical dashed lines indicate the frequency intervals $\Delta_\mathrm{low}$ (800-7500~cm$^{-1}$) and $\Delta_\mathrm{high}$ (7500-12500~cm$^{-1}$). (b) Effective number of electrons in the frequency intervals $\Delta_\mathrm{low}$ and $\Delta_\mathrm{high}$ as a function of pressure, showing the spectral weight transfer from high to low energies with increasing pressure. The vertical grey shaded areas mark the critical pressure ranges of the phase transitions as explained in the text.}\label{figure3}
\end{figure}

To quantify the spectral weight transfer related to interband transitions, we subtracted the Drude component from the total optical conductivity for each pressure to obtain the interband contributions to the optical conductivity, $\sigma_{1,\mathrm{inter}}$. Afterwards, we calculated the difference spectra of the interband transitions $\Delta\sigma_{1,\mathrm{inter}}$ according to $\Delta\sigma_{1,\mathrm{inter}}(\omega,P)=\sigma_{1,\mathrm{inter}}(\omega,P)-\sigma_{1,\mathrm{inter}}(\omega,1.2~\mathrm{GPa})$. We show the contour plot of $\Delta\sigma_{1,\mathrm{inter}}$ in Fig.~\ref{figure3}(a). Next, we determined the effective number of electrons $N_\mathrm{eff}$ per atom contributing to the interband optical conductivity from the spectral weight analysis, applying the sum rule:
\begin{equation}
N_\mathrm{eff}(\omega_\mathrm{l},\omega_\mathrm{u})=\frac{2m_0}{\pi e^2N}\int_{\omega_\mathrm{l}}^{\omega_\mathrm{u}}\sigma_{1,\mathrm{inter}}(\omega)d\omega \quad ,
\end{equation}
where $\omega_\mathrm{l}$ and $\omega_\mathrm{u}$ are the lower and upper limits of the frequency range, respectively, $m_0$ is the free electron rest mass, and $N$=12 is the number of atoms per unit cell \cite{Dressel.2002}. We chose the two frequency intervals $\Delta_\mathrm{low}$ (800-7500~cm$^{-1}$) and $\Delta_\mathrm{high}$ (7500-12500~cm$^{-1}$) for our calculations, since the low-frequency interval corresponds to the area around the contribution L2 at $\sim$4500~cm$^{-1}$ and below, and the high-frequency interval mainly covers the range of the higher energy excitations L4/L5 at $\sim$10000~cm$^{-1}$. The pressure dependence of the effective number of electrons, corresponding to these two frequency ranges, is plotted in Fig.~\ref{figure3}(b). While $N_\mathrm{eff}$ for $\Delta_\mathrm{low}$ is increasing with increasing pressure, $N_\mathrm{eff}$ for $\Delta_\mathrm{high}$ is decreasing by approximately the same amount. Since $\sigma_1(\omega_\mathrm{center})$, i.e., in the region between $\Delta_\mathrm{low}$ and $\Delta_\mathrm{high}$, stays constant throughout all pressures, this behavior rules out a simple shifting of bands.
Further bandstructure-related and band-selective optical conductivity calculations are needed in order to identify the pressure-induced changes in the electronic bandstructure.

\begin{figure}[t]
\includegraphics[width=0.45\textwidth]{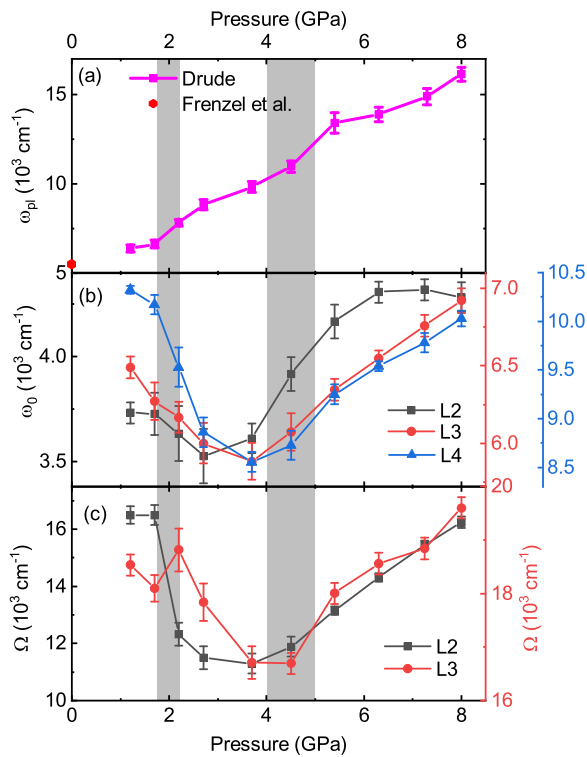}
\caption{(a) Plasma frequency of the Drude component, $\omega_\mathrm{pl,D}$, as a function of pressure. The value of $\omega_\mathrm{pl,D}$ at ambient pressure was extracted Frenzel et al.\ \cite{Frenzel.2017} as the average of the two measured polarization directions. (b) Resonance frequency $\omega_{0,j}$ of the Lorentz contributions L2 (black square), L3 (red dot), and L4 (blue triangle) as a function of pressure. (c) Oscillator strength, as given by $\Omega_j$ in Equ.\ (\ref{equ:Drude-Lorentz}), for the Lorentz contributions L2 (black square) and L3 (red dot) as a function of pressure. The vertical grey shaded areas mark the critical pressure ranges of the phase transitions as explained in the text.}\label{figure4}
\end{figure}

In addition to the pronounced spectral weight transfer between the interband transitions, we observe a peculiar pressure behavior of the plasma frequency of the Drude component $\omega_\mathrm{pl,D}$, as extracted from the fittings, which is plotted in Fig.~\ref{figure4}(a). The plasma frequency is linked to the carrier density $n$ according to $\omega_\mathrm{pl,D}^2=4\pi n e^2/m^*$, with the electron charge $e$ and the effective mass $m^*$. At 1.2~GPa, $\omega_\mathrm{pl,D}$ amounts to 6396$\pm$205~cm$^{-1}$, which matches well with literature data at ambient conditions \cite{Frenzel.2017}. The pressure dependence of $\omega_\mathrm{pl,D}$ shows two anomalies:
 Above 2~GPa, the $\omega_\mathrm{pl,D}$ increases significantly, consistent with the increasing metallicity of WTe$_2$ under pressure \cite{Cai.2015,Pan.2015}. Besides the onset of the pressure-induced increase at $\sim$2~GPa, there seems to be a second anomaly at $\sim$4.5~GPa in the pressure dependence of $\omega_\mathrm{pl,D}$, namely a small step with a decrease of the slope of the curve.

Anomalies in the pressure dependence are also found for other fitting parameters. Around 2~GPa, the resonance frequency $\omega_0$ as well as the oscillator strength $\Omega_j$ of the low-energy Lorentz contributions L2 and L3 show a sudden drop and at $\sim$4.5~GPa start to increase again as depicted in Fig.~\ref{figure4}(b) and (c), respectively. Furthermore, at $\sim$2~GPa the contribution L4 splits into
two components, L4 and L5 (see grey arrows in Fig.~\ref{figure2}). This splitting is responsible for the drop of the resonance frequency $\omega_0$ of L4 in Fig.\ref{figure4}(b), and at $\sim$4.5~GPa there is an additional inflection point, similar to L2 and L3.
Besides, several Lorentz oscillators show anomalous behavior in their scattering rate at 2 and 4.5~GPa (Fig.~\ref{suppl3} in the appendix).

\section{DISCUSSION}

According to electronic band structure calculations for T$_\mathrm{d}$-WTe$_2$ under pressure by Pan et al. \cite{Pan.2015},
we attribute the observed increasing metallicity stemming to the increasing size of the electron and hole pockets, and thus to an increasing density of states at the Fermi level $N(E_\mathrm{F})$ upon pressure application. Lu et al. \cite{Lu.2016} confirmed this behavior by calculations for 1T'-WTe$_2$. The band structure and Fermi surface of the 1T' phase were stated to be very similar to the T$_\mathrm{d}$ structure. Furthermore, a Lifshitz transition, i.e., a change in the shape of the Fermi surface \cite{Lifshitz.1960}, was found for pressures below 5~GPa \cite{Lu.2016}. In fact, Cai et al. \cite{Cai.2015} experimentally observed a change of the Fermi surface topology already at $\approx$1~GPa. Since the shape of the Fermi surface directly influences the optical properties, we should see signatures of the Lifshitz transition in our optical data.
Indeed, besides the onset of the increase of $\omega_\mathrm{pl,D}$, the pronounced high-energy Lorentz contribution L4 splits into two components L4 and L5, of which the former shifts abruptly to lower energies between 1.7 and 2.2~GPa, whereas the latter shifts to higher energies upon further pressure increase.  This manifests itself in the fact that we have to use slightly different fitting models to describe the spectra at low ($<$ 2~GPa) and at high ($>$ 2~GPa) pressures (see Fig.~\ref{figure2} and Fig.~\ref{suppl2} and Fig.~\ref{suppl3} in the appendix). Such a splitting of an absorption band is a strong indication for a phase transition, consistent with the observation of an electronic phase transition reported in Ref.\ \cite{Cai.2015}. Furthermore, our findings are consistent with the \textit{ab initio} calculations of Refs.\ \cite{Pan.2015,Lu.2016}, where they observe that several electronic bands are pushed towards $E_\mathrm{F}$ and cross it between 0 and 5~GPa without substantially changing the slope of the bands.
Such changes in the bandstructure are further supported by an anomaly at $\sim$2~GPa in the resonance frequency, oscillator strength, and scattering rate for several Lorentz contributions, as described above.

Furthermore, several Lorentz contributions show at $\sim$4.5 GPa an inflection in their resonance frequency and oscillator strength, as well as a broadening. This inflection coincides with the aforementioned anomaly of $\omega_\mathrm{pl,D}$ at around 4.5~GPa.
We relate this second anomaly to the pressure-induced structural phase transition from T$_\mathrm{d}$- to 1T'-WTe$_2$ reported in the literature:
According to pressure-dependent XRD, Raman spectroscopy, and electrical transport measurements a T$_\mathrm{d}$ to 1T' transition with a broad transition range from 6.0 to 15.5~GPa occurs at room temperature, and superconductivity was observed in both polytypes \cite{Zhou.2016} at low temperature. Lu et al. \cite{Lu.2016} identified a similar structural transition between $\sim$4 and 11~GPa at room temperature by XRD and Raman spectroscopy supported by \textit{ab initio} calculations, proposing superconductivity to emerge from the high-pressure 1T' phase. Xia et al. \cite{Xia.2017} also found such a structural phase transition at 8-10~GPa at room temperature in the plane-vertical Raman response, linking it to the emerging superconductivity \cite{Kang.2015}\footnote{One has to note the different pressures at which superconductivity has been observed, namely 2.5 (no PTM) \cite{Pan.2015}, 4.0 (PTM: Daphne 7373) \cite{Zhou.2016} and 10.5~GPa (PTM: NaCl) \cite{Kang.2015}, as well as the different pressure ranges of the structural phase transition, namely 6-15.5 (PTM: Daphne 7373) \cite{Zhou.2016}, 4-11 (PTM: Ar, Ne) \cite{Lu.2016} and 8-10~GPa (PTM: methanol-ethanol 4:1 mixture) \cite{Xia.2017}. The discrepancies have not been clarified but might be related to the different pressure transmitting media used. Additionally, the crystal quality might have an influence on the electronic properties of WTe$_2$ \cite{Ali.2015}.}.
We note here that in the XRD study of Kang et al. \cite{Kang.2015} no signs for a structural phase transition were found under pressure, and therefore a Lifshitz transition at 10.5~GPa was proposed to be responsible for the emerging superconductivity.

According to the similarity of the electronic band structures of T$_\mathrm{d}$- and 1T'-WTe$_2$, we don't expect to see a strong signature of the pressure-induced structural phase transition in the pressure-dependent optical conductivity. However, the anomalies in Fig.~\ref{figure4} are significant. Our data therefore support the occurrence of a phase transition at $\sim$4.5~GPa, coinciding with the reported structural phase transition from T$_\mathrm{d}$- to 1T'-WTe$_2$ \cite{Zhou.2016,Lu.2016,Xia.2017}. As mentioned above, this phase transition is reported to occur across a rather broad pressure range.

\section{CONCLUSION}
According to the pressure dependence of the optical conductivity of WTe$_2$, we observe a large transfer of spectral weight from high to low energies induced by external pressure. This finding points to strong changes in the electronic bandstructure of WTe$_2$ under pressure. Furthermore, there are indications for two pressure-induced phase transitions at around 2 and 4.5~GPa according to the pressure dependence of several optical parameters. The first transition most probably is of electronic type, i.e., it substantially affects the electronic bands, with a strong increase in the Drude plasma frequency. The transition at $\sim$4.5~GPa corresponds to the structural phase transition from the T$_\mathrm{d}$- to the 1T' phase, where several optical parameters show an anomaly in their pressure evolution. Overall, our findings show the profound sensitivity of the bandstructure of WTe$_2$ to external pressure.



\section*{Appendix}

\begin{figure*}[t]
\includegraphics{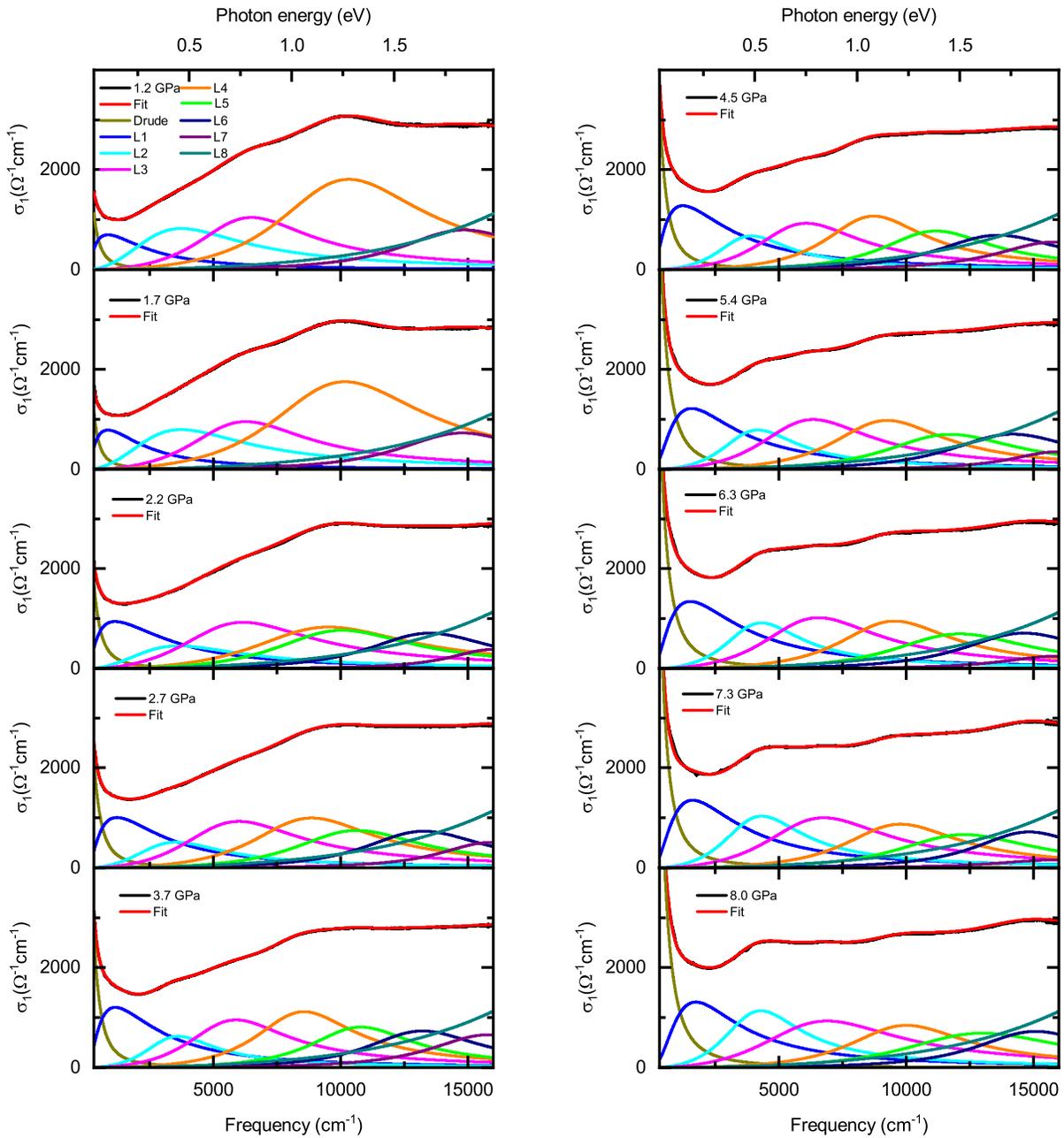}
\caption{Contributions to the pressure-dependent optical conductivity $\sigma_1$ of WTe$_2$, as obtained from Drude-Lorentz fitting.
The contribution L8 was kept constant for all fittings, as it is part of the high-energy extrapolation.}\label{suppl2}
\end{figure*}

\begin{figure*}[t]
\includegraphics[width=0.9\textwidth]{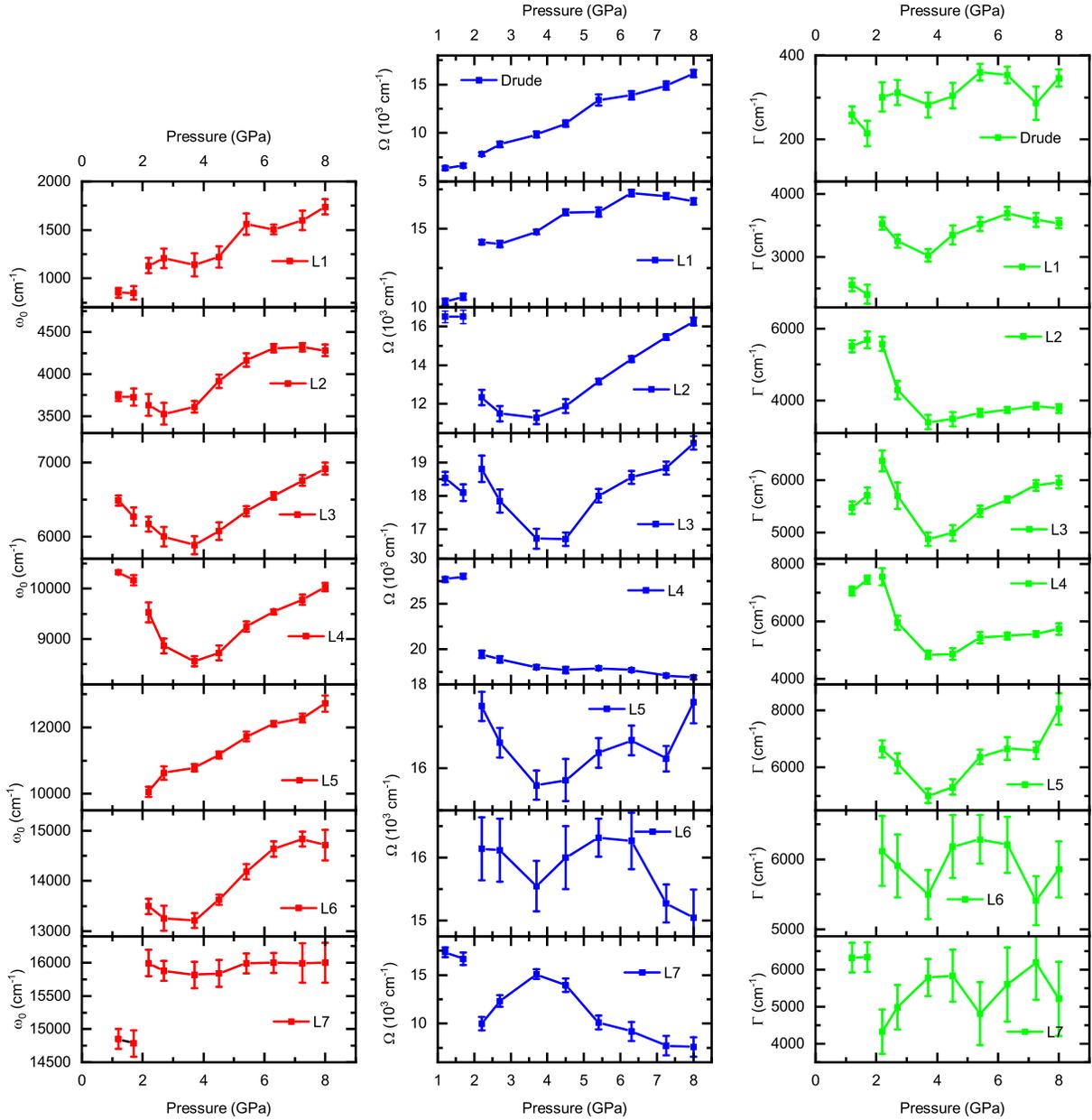}
\caption{Pressure-dependence of $\omega_0$, $\omega_\mathrm{pl,D}$, $\Omega$, and $\Gamma$ of the various contributions of the Drude-Lorentz fittings shown in Fig.~\ref{suppl2} and described in the main text.}\label{suppl3}
\end{figure*}

In the appendix we show additional optical data, namely the optical conductivity $\sigma_1(\omega)$, including the Drude-Lorentz fits and contributions, for all pressures (see Fig.~\ref{suppl2}), and the pressure evolution of all the parameters obtained from these fits (see Fig.~\ref{suppl3}).

\end{document}